\documentclass[aps,prl,twocolumn,showpacs,preprintnumbers,amsmath,amssymb,superscriptaddress]{revtex4-1}

\usepackage[T1]{fontenc}
\usepackage[latin9]{inputenc}
\usepackage{graphicx}
\usepackage{epstopdf}

\renewcommand{\vec}{\mathbf}
\renewcommand{\v}{\vec}
\renewcommand{\d}{\mathrm{\,d}}

\renewcommand{\theta}{\vartheta}

\begin{document}

\title{Rydberg-induced Solitons: Three-dimensional Self-trapping of Matter Waves}

\author{F. Maucher}
\affiliation{Max Planck Institute for the Physics of Complex Systems, 01187 Dresden,
Germany}
\author{N. Henkel}
\affiliation{Max Planck Institute for the Physics of Complex Systems, 01187 Dresden,
Germany}
\author{M. Saffman}
\affiliation{Department of Physics, University of Wisconsin, Madison, Wisconsin 53706, USA}
\author{W. Kr\'olikowski}
\affiliation{Laser Physics Centre, Research School of
Physics and Engineering, Australian National University, Canberra, ACT
0200, Australia
}
\author{S. Skupin}
\affiliation{Max Planck Institute for the Physics of Complex Systems, 01187 Dresden,
Germany}
\affiliation{Friedrich Schiller University, Institute of Condensed Matter Theory and
Optics, 07743 Jena, Germany}
\author{T. Pohl}
\affiliation{Max Planck Institute for the Physics of Complex Systems, 01187 Dresden,
Germany}

\begin{abstract}
We propose a scheme for the creation of stable three-dimensional bright solitons in Bose-Einstein condensates, i.e., the matter-wave analog of so-called spatio-temporal "light bullets".
Off-resonant dressing to Rydberg $nD$-states is shown to provide nonlocal attractive interactions, leading to self-trapping of mesoscopic atomic clouds by a collective excitation of a Rydberg atom pair. We present detailed potential calculations, and demonstrate the existence of stable solitons under realistic experimental conditions by means of numerical simulations.
\end{abstract}


\maketitle

Self-trapped nonlinear waves and the possibility to create ''particle-like'' wave packets have fascinated scientists
over the last decades ~\cite{Chiao:64:PRL,Stegeman:99:Science,Strecker:02:nature,malomed05}.
In nonlinear optics, the creation of stable three-dimensional bright solitons $-$ so called ''light-bullets`` \cite{Silberberg:90:OptLett} $-$ has been under active pursuit \cite{malomed05}, but was realized only recently in a discrete setting of waveguide arrays~\cite{Minardi:11:PRL}.
One major obstacle stems from the fact that nonlinear confinement usually comes hand in hand with collapse instabilities~\cite{Berge:pr:303:259}.
In principle, this problem can be overcome via nonlocal nonlinearities, where the nonlinear self-induced potential at a particular point
in space depends also on the nearby wave amplitudes.
While experiments have established two-dimensional optical solitons in a continuous nonlocal medium (see e.g.~\cite{Conti:prl:92:113902}), the realization of their three-dimensional counterparts remains an elusive goal.
On the other hand, Bose-Einstein condensates (BECs) have emerged as clean model systems to investigate soliton formation~\cite{Khaykovich2002,Strecker:02:nature,Gerton:np:2000}, and, more recently, to study the effects of nonlocal nonlinearities, as arising from dipolar interactions~\cite{lahaye}. Such interactions can give rise to stable two-dimensional bright solitons \cite{pedri2005}, while the partially attractive nature of dipolar interactions precludes stability of three-dimensional solitons due to collapse~\cite{laha08}.

\begin{figure}
\includegraphics[width=0.99\columnwidth]{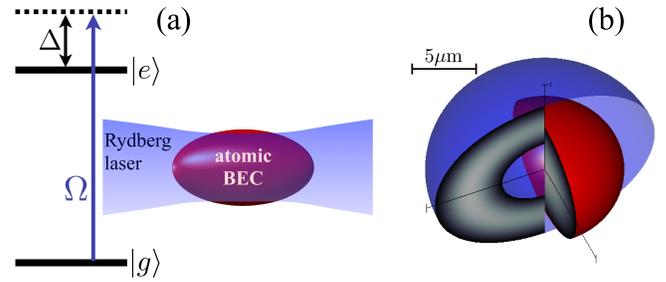}
\caption{(a) Simplified level scheme of the considered Rydberg-dressing approach. The off-resonant coupling of condensed ground state ($|g\rangle$) atoms to strongly interacting Rydberg states ($|e\rangle$) gives rise to  a stable bright soliton bound by a rim of a collective quantum excitation of a Rydberg atom pair, shown in (b).
Depicted is a numerically obtained soliton formed by dressing a Rb condensate to $65D_{3/2}$ Rydberg states with a Rabi frequency $\Omega/2\pi=0.5$MHz and laser detuning $\Delta/2\pi=32$MHz. Shown are half-maximum iso-density surfaces of the BEC (inner red sphere) and the Rydberg-pair density (outer blue shell), while the gray-scale gives the interior density.}
\label{fig1}
\end{figure}

In this work, we propose an experimental scheme for the creation of
''matter wave bullets'', i.e., stable three-dimensional (3D) bright
solitons in a Bose-Einstein condensate.
The approach is based on optical dressing of a groundstate atom BEC to highly excited Rydberg states~\cite{Henkel:PRL,pupillo10,Cinti10} (see Fig.~\ref{fig1}). Within detailed potential calculations for Rubidium condensates we identify an appropriate range of Rydberg states that provides attractive nonlocal nonlinearities, which do not lead to condensate collapse. We present extensive 3D simulations of the underlying Gross-Pitaevskii equation that demonstrate the emergence of stable condensate bullets for realistic experimental parameters and scenarios of their creation. The calculations show that mesoscopic numbers of ultracold atoms can be bound together by a single pair of strongly interacting Rydberg atoms which is coherently shared throughout the condensate (see Fig.~\ref{fig1}). This exotic state $-$ arising from intricate coupling between internal and translational atomic quantum dynamics $-$ is shown to live over hundreds of milliseconds and could, thus, provide the first realization of 3D bright solitons in atomic BECs.

\begin{figure}
\includegraphics[width=0.9\columnwidth]{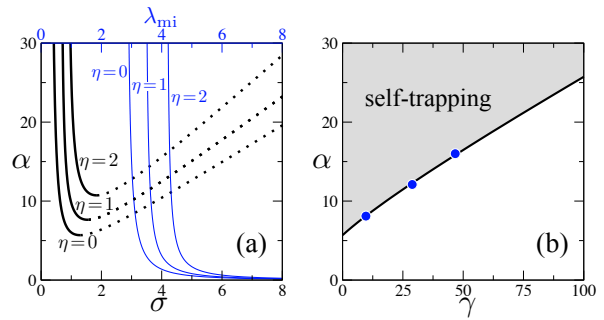}
\caption{(a) Existence curves for stable (thick solid lines) and unstable (dotted lines) solitons for several $\eta=\gamma/\alpha$ and typical length scales of modulational instability (thin blue lines) (b) Corresponding parameter regions for self-trapping. The dots show numerical results using the potential Eq.~(\ref{pot_real}) for a Rb condensate with atom numbers $N=1000\,,\:3000\,,\:4850$.}
\label{fig:phasediagram}
\end{figure}

In the zero temperature limit the BEC wavefunction $\psi$ obeys the Gross-Pitaevskii equation (GPE)
\begin{equation}
i\hbar\partial_t \psi = \left[-\frac{\hbar^2\nabla^2}{2M}+ g|\psi|^2 + \int W({\bf r}-{\bf r}^{\prime}) |\psi({\bf r}^{\prime})|^2\d^3{\bf r}^{\prime}\right]\psi,
\label{GPE}
\end{equation}
where $N=\langle\psi|\psi\rangle$ is the number of atoms, $M$ denotes the atomic mass and $g=4\pi \hbar^2a/M$ describes the strength of local interactions due to s-wave scattering with scattering length $a$. The range of the potential $W$ greatly exceeds the average interatomic distance, and, thus gives rise to the nonlocal nonlinearity in Eq.~(\ref{GPE}). Let us first consider a simplified, spherically symmetric potential of the form
\begin{equation}\label{potential}
W({\bf r}-{\bf r}^{\prime})=\frac{\tilde{C}_6}{R_{\rm c}^6+|{\bf r}-{\bf r}^{\prime}|^6}\;.
\end{equation}
Typical values of the parameters $\tilde{C}_6<0$ and $R_{\rm c}$ will be discussed below in our explicit calculations of the dressing-induced interaction. As we will see, this generic potential readily captures the essential physics of soliton formation.
Introducing appropriate length ($R_{\rm c}$) and time ($\tau=R_{\rm c}^2M/\hbar$) scales, Eq.~(\ref{GPE}) can be re-expressed as
\begin{equation}
i\partial_t \psi = \left[-\frac{\nabla^2}{2}+ \gamma|\psi|^2 - \alpha\int \frac{|\psi({\bf r}^{\prime})|^2\d^3{\bf r}^{\prime}}{1+|{\bf r}-{\bf r}^{\prime} |^6}\right]\psi.
\label{GPE_scaled}
\end{equation}
This simple scaling shows that the BEC properties are solely determined by two dimensionless parameters $\alpha=-MN\tilde{C}_6/\hbar^2R_c^4$ and $\gamma=4\pi a N/R_{\rm c}$, where $\langle\psi|\psi\rangle =1$.
In order to find approximate soliton solutions we first use a Gaussian trial wavefunction
of width $\sigma$. Following standard variational analysis we find an implicit relation
\begin{equation}\label{eq:A_of_sigma}
\alpha=\frac{1}{2}\frac{24\pi^{3}\sigma^4}{24Z - \left( 3\eta\sigma^3\pi^{3/2}\sqrt{2}+4\sigma \partial_\sigma Z \right)},
\end{equation}
involving the interaction ratio $\eta = \gamma/\alpha$ and the width of condensate $\sigma$.
The integral $Z = -\int W({\bf r} - {\bf r}^\prime) e^{-{\bf r}^2/\sigma^2} e^{-{\bf r}^{\prime 2}/\sigma^2} \mathrm{d}\bf{r}^\prime\mathrm{d}\bf{r}$ can be evaluated analytically in terms of hypergeometric functions. In Fig.~\ref{fig:phasediagram}(a) we show the resulting existence curves for different values of the interaction ratio $\eta$. As expected, the condensate size is on the order of $R_{\rm c}$ and increases with increasing local repulsion. According to the Vakhitov-Kolokolov criterion~\cite{Kolokolov:RPQE:1974}, the stability of the solitons is expected to change at the minima of the existence curves. This is confirmed by our  numerical solutions of Eq.~(\ref{GPE_scaled}), which show that solitons on the left branches are stable, whereas the right branches are unstable. Consequently, the minima mark the critical $\alpha$ above which stable self-trapped solitons can exist. The corresponding stability diagram is shown in Fig.~\ref{fig:phasediagram}(b). For vanishing local repulsion ($\gamma=0$) this critical value is found to be $\alpha=5.7$, and it increases almost linearly with $\gamma$ in the considered parameter range.

While this analysis yields a simple picture for the stability conditions of self-trapped BECs, their dynamical formation deserves some additional discussion.
In particular, we note that attractive kernels generally lead to long wavelength modulation instabilities, i.e an exponential growth of periodic perturbations on a constant density background.
The wavenumber $k_{\rm mi}$ of the periodic perturbation with maximum growth-rate gives the typical length scale $\lambda_{\rm mi}=2\pi k_{\rm mi}^{-1}$ of the unstable modulation. As shown in Fig.~\ref{fig:phasediagram}(a), the soliton width $\sigma$ of the stable branch is significantly smaller than $k_{\rm mi}$. Hence, there is a wide range of initial condensate states with diameters smaller than $\lambda_{\rm mi}$ and with sufficiently large values of $\alpha$ to eventually form a stable soliton.

Having established the general conditions for self-trapping based on the simplified potential Eq.~(\ref{potential}), we now discuss its physical
realization via off-resonant dressing of ground state atoms with high-lying Rydberg states. Let us consider first the most simple case of
a single non-degenerate Rydberg state coupled to the atomic groundstate with a Rabi frequency $\Omega/2\pi$ and laser detuning $\Delta/2\pi$ [see Fig.~\ref{fig1}(a)].
This admixes a small fraction $\nu=(\Omega/2\Delta)^2$ of Rydberg character into the atomic ground states.
For two distant atoms, this leads to an effective interaction $\nu^2C_6/r^6$, arising from the strong van der Waals interaction $C_6/r^6$
between Rydberg atoms. At smaller distances $r$ the doubly excited state is blocked by the van der Waals shift \cite{jaksch00}, such that the
effective interaction saturates. Altogether, this yields an effective potential of the type Eq.~(\ref{potential}), where $\tilde{C}_6=\nu^2C_6$
and $R_{\rm c}=(-C_6/2\hbar\Delta)^{1/6}$~\cite{Henkel:PRL}.

\begin{figure}
\includegraphics[width=0.99\columnwidth]{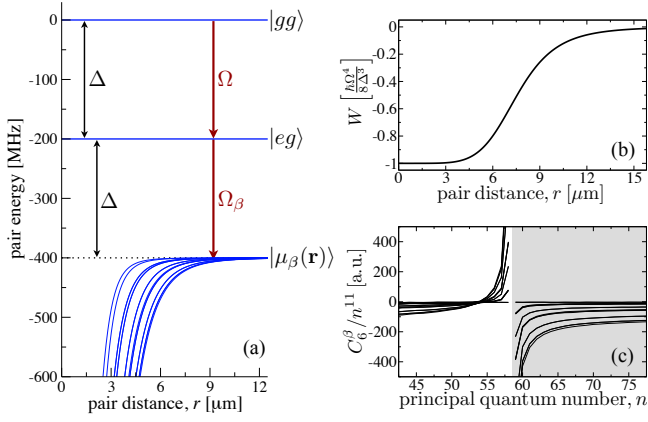}
\caption{(a) Two atom level-scheme for dressing to $65D_{3/2}$ Rydberg states and Rydberg interactions $V^{\rm{Ryd}}_\beta (r)$. (b) Resulting single potential curve ($\tilde{m}_J=3/2$) for $\vartheta=0$ according to Eq.~(\ref{pot_real}).
(c) Attractive and repulsive branches of interaction strengths $C^{(\beta)}_6$ vs.\ $n$. For $n\ge59$, all curves are attractive.
\label{fig3}}
\end{figure}

This scenario may be realized with $nS_{1/2}$ Rydberg states of Rubidium atoms. However, the van der Waals interactions of alkaline atoms are entirely repulsive for highly excited $S$-states, such that the above scheme only produces repulsive nonlinearities. We, therefore, resort to higher angular momentum
($nP_{J}$ or $nD_{J}$) states, accessible either by single- or two-photon transitions.
The underlying Rydberg-Rydberg atom interactions are calculated by diagonalizing the two-atom Hamiltonian, which includes the dipole-dipole coupling  between Rydberg states.
Since there are ($2J+1$) Zeeman degenerate Rydberg states ($m_J=-J,-J+1,...,J$) one obtains a set of $(2J+1)^2$ potential curves
$V_{\rm ryd}^{(\beta)}({r})$ [see Fig.~\ref{fig3}(a)] with respective molecular eigenstates $|\mu_{\beta}({\bf r})\rangle$ \cite{walker08}. 
For distances $r$ 
larger than the van der Waals radius $R_{\rm vdW}$ \cite{walker08} the potentials behave like $V_{\rm ryd}^{(\beta)}({r})\sim C_6^{(\beta)}/r^6$. The eigenstates depend
on the interatomic distance $r$ and on the angle $\theta$ between the molecular axis ${\bf r}$ and the quantization axis defined by
the coupling laser.
Note that all of the potential curves $V_{\rm ryd}^{(\beta)}({r})$ must be of negative sign. Already a single repulsive curve would lead to resonant, simultaneous excitation of two Rydberg atoms at a distance for which $V_{\rm ryd}^{(\beta)}({r})=2\hbar\Delta$,
and cause rapid loss on a timescale $\sim100\mu$s for typical Rydberg states (see below).
This requirement significantly restricts the range of appropriate principal quantum numbers $n$ and
angular momentum states. We have performed potential calculations for different Rydberg states of Rubidium, and found $nD_{3/2}$-interactions to be most favorable, as they yield entirely attractive curves for $n\geq 59$ [cf.\ Fig.~\ref{fig3}(c)].

\begin{figure}
\includegraphics[width=0.99\columnwidth]{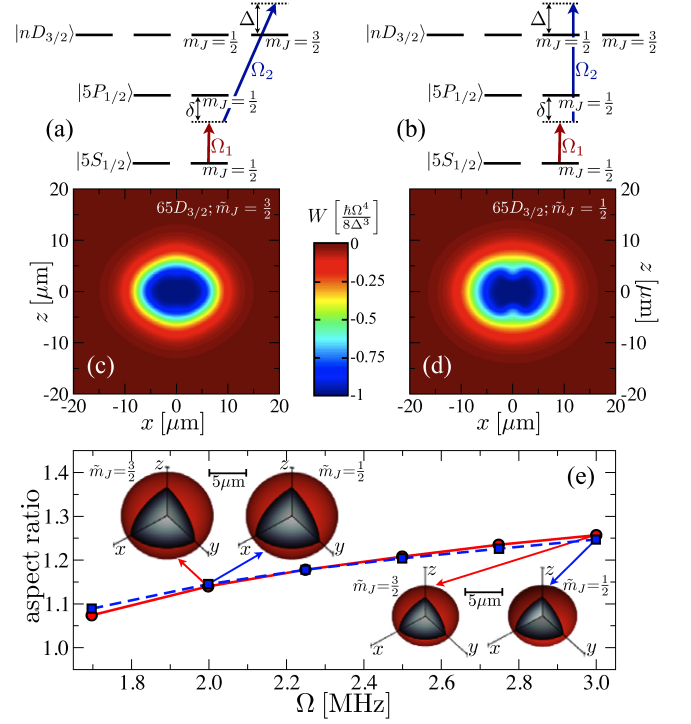}
\caption{Effective ground state potentials (c,d) resulting from two-photon $nD_{3/2}$ state dressing (a,b) to $\tilde{m}_J=3/2$ (a,c) and $\tilde{m}_J=1/2$ (b,d) Rydberg states, with a two-photon Rabi frequency $\Omega/2\pi=(\Omega_1\Omega_2/2\delta)/2\pi=0.5$MHz and laser detuning $\Delta/2\pi=32$MHz. The aspect ratio of the resulting soliton is shown in (e) as a function of $\Omega$ for $\tilde{m}_J=3/2$ (circles) and $\tilde{m}_J=1/2$ (squares). The insets in (e) show respective 3D soliton profiles [cf.\ Fig.~\ref{fig1}(b)] for $\Omega/2\pi=0.3$MHz and $0.5$MHz.
\label{fig4}}
\end{figure}

The corresponding two-atom level scheme for dressing to such Rydberg states is shown in Fig.~\ref{fig3}(a). Upon proper choice of laser polarization,
the atomic ground state ($|g\rangle\equiv|5S_{1/2}\rangle$) can be selectively coupled to a single $m_J$-state out of the degenerate
$nD_{3/2}$-manifold [cf.\ Fig.~\ref{fig4}(a,b)], denoted by $|e\rangle\equiv|nD_{3/2}(\tilde{m}_J)\rangle$. Generally, the singly excited two atom states ($|ge\rangle$, $|eg\rangle$)
are, however, coupled to all molecular eigenstates $|\mu_{\beta}({\bf r})\rangle$, due to $m$-mixing by the dipole-dipole interaction $\hat D({\bf r})$.
The corresponding Rabi frequencies $\Omega_{\beta}=\langle ee|\mu_{\beta}({\bf r})\rangle\Omega$ are given by the overlap with the laser-coupled Rydberg states, and thus depend on the interatomic separation vector.

Assuming $\Omega\ll\Delta$ the corresponding two-atom Hamiltonian
can be diagonalized within fourth order perturbation theory. Omitting terms that do not depend on ${\bf r}$, one obtains a single anisotropic effective interaction potential
\begin{equation}\label{pot_real}
W(\v r) = \frac{\hbar\Omega^4}{8\Delta^3}\sum_{\beta=1}^{(2J+1)^2}|\langle ee|\mu_{\beta}(\v r)\rangle|^2\frac{V_{\rm ryd}^{(\beta)}(r)}{{2\hbar\Delta}-V_{\rm ryd}^{(\beta)}(r)}
\end{equation}
between two dressed ground state atoms. This two-atom calculation can be straightforwardly extended to the $N$-atom case.
As long as the total number of excited Rydberg atoms $N_{\rm ryd}=\left(\tfrac{\Omega}{2\Delta}\right)^2N\ll1$ \cite{Honer:prl:105:160404}, this simply yields $\sum_{i<j}W(|{\bf r}_i-{\bf r}_j|)$ for the total potential energy of $N$ atoms at positions ${\bf r}_i$~\cite{Henkel:PRL}. The mean field dynamics of the BEC is, thus, governed by the GPE~(\ref{GPE}) and Eq.~(\ref{pot_real}).

To further illuminate the physical mechanism behind the observed self-trapping, let us consider the resulting many-body states of the Rydberg-dressed condensate. From the described perturbation theory we find that the second order contribution to the many-body groundstate
\begin{equation}\label{correlation}
|G^{(2)}\rangle=\frac{\Omega}{2\Delta}\sum_{i<j}\sum_{\beta}\frac{\hbar\Omega_{\beta}}{2\hbar\Delta -V^{(\beta)}_{\rm ryd}(r_{ij})}|\mu_{\beta}({\bf r}_{ij})\rangle\prod_i\psi({\bf r}_i)
\end{equation}
is a coherent superposition of Rydberg atom pair states, whose correlation function drops to zero within a distance for which $V^{(\beta)}_{\rm ryd}(r)=-2\hbar\Delta$. This radius coincides with the typical size of the self-trapped BEC, such that two atoms are excited on opposite sides of the soliton. According to Eq.~(\ref{correlation}), these pairs are coherently shared among all atoms, resulting in a collective shell of Rydberg excitations [c.f. Fig.\ref{fig1}(b)]. It is the strong attraction between these Rydberg atom pairs that self-confines the BEC.

In the following, we present explicit calculations for a $^{87}$Rb condensate of $N=1000$ atoms dressed to $65D_{3/2}$ Rydberg states with $\Delta/2\pi=32$MHz, based on a 3D integration of Eqs.~(\ref{GPE}) and (\ref{pot_real}). To ensure applicability of our potential calculations, we consider Rabi frequencies between $1$MHz and $3$MHz for which $0.006 \le N_{\rm ryd} \le 0.06$.
Figs.\ref{fig4}(c,d) show that due to the angular dependence of the underlying Rydberg wave functions the resulting interaction potentials are anisotropic [in contrast to our isotropic model potential Eq.~(\ref{potential})], with a cylindrical symmetry around the quantization axis ($z$-axis). Consequently, the resulting self-trapped ground states shown in Fig.~\ref{fig4}(e) are also slightly asymmetric. However, the kinetic pressure as well as the local contact interaction both tend to round out the soliton solutions, such that our previous analysis using the simplified potential Eq.~(\ref{potential}) is expected to remain valid. In order to make quantitative comparisons, we define an angular dependent radius $R_{\rm c}(\theta)$ at which the potential Eq.~(\ref{pot_real}) assumes half its minimum value $W(0)=-\hbar\Omega^4/8\Delta^3$. For the chosen parameters this radius is considerable larger than $R_{\rm vdW}\approx 3\mu$m, such that the actual potential Eq.(\ref{pot_real}) closely resembles our model potential Eq.~(\ref{potential}). Using the average softcore radius $\bar{R}_{\rm c}=\pi^{-1}\int_0^{\pi} R_c(\theta) \sin{\theta} {\rm d}\theta$ as a natural length scale and $\bar{\tau}=\bar{R}_{\rm c}^2M/\hbar$ to scale times, we obtain a dimensionless GPE of the type of Eq.~(\ref{GPE_scaled}), with $\alpha=N\Omega^4\bar{R}_{\rm c}^2M/8\hbar\Delta^3$ and $\gamma=4\pi aN/\bar{R}_{\rm c}$. The numerically obtained critical interaction strengths for soliton existence, shown in Fig.~\ref{fig:phasediagram}(b)
are in excellent agreement with the variational results based on the simple model potential Eq.~(\ref{potential}).

\begin{figure}
\includegraphics[width=0.8\columnwidth]{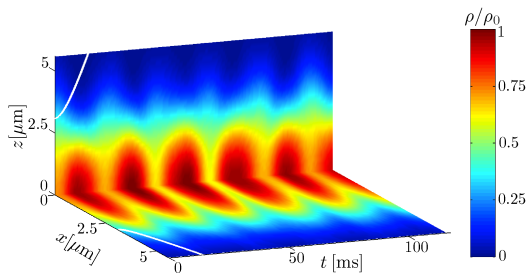}
\caption{Self-trapped condensate dynamics after instantaneous turn-on of the dressing lasers and sudden release from a harmonic trap ($\omega_{\rm tr}=170$Hz). The white lines give the width of a freely expanding BEC.
$65D_{3/2}$ Rydberg states are coupled with $\Omega/2\pi=0.5$MHZ and $\Delta/2\pi=32$MHz.
\label{fig5}}
\end{figure}

Experimentally, coherent Rydberg excitation of cold atoms has been demonstrated under various conditions \cite{RydExp}.
Ideally, the preparation would start from a 
cylindrically symmetric trap to initialize a BEC already close to the solitary ground state. However, in order to demonstrate robustness of the soliton, Fig.~\ref{fig5} shows the dynamics of a significantly different initial condensate after sudden release from a spherical harmonic trap and simultaneous turn-on of the dressing lasers. Compared to the undressed BEC expansion, self-trapping becomes evident within $10$~ms, manifested in stable oscillations of the condensate size. On the other hand, the lifetime of the dressed BEC is as long as several $100$~ms, being predominantly limited by the decay of the weakly admixed $6P_{1/2}$ state \cite{Henkel:PRL} [cf.\ Figs.\ref{fig4}(a,b)]. A verification of the predicted self-trapped "matter wave bullets", thus, appears to be well in reach of current experimental capabilities.

In conclusion, we have shown that off-resonant dressing of BECs to attractively interacting Rydberg states provides a promising route for the first realization of stable self-trapped three-dimensional bright solitons. While we chose  Rb($nD_{3/2}$) atoms as one relevant example, the proposed scheme generally applies to Rydberg states of any atomic species with sign-definite attractive van der Waals interactions. Explicit potential calculations and numerical simulations of the resulting GPE have been performed to demonstrate experimental feasibility and to work out appropriate parameters. In addition, we have shown that a simple isotropic model potential captures the essential physics of the observed soliton formation, which may be useful for future theoretical studies on, e.g., 3D soliton interaction or higher-order states \cite{Mihalache:02:PRL}. The anisotropy of the interaction together with its tunability may also open up new routes to transfer angular momentum to the condensate \cite{Tikhonenkov:PRL:2008}.

\end{document}